Student Research Project
by
Aaron Grunthal

# Efficient Indexing of the BitTorrent Distributed Hash Table

University of Applied Sciences Esslingen

**Abstract.** The following paper presents various methods and implementation techniques used to harvest metadata efficiently from a Kademlia Distributed Hashtable (DHT) as used in the BitTorrent P2P network to build an index of publicly available files in the BitTorrent ecosystem.

The indexer design makes various tradeoffs between throughput and fairness towards other DHT nodes while also being scaleable in a distributed environment.

Advisor:     Prof. Dominik Schoop
Faculty:     Information Technology

# Table of Contents

# 1 Introduction

The BitTorrent [1] P2P filesharing ecosystem consists of several core components, each fulfilling specialized roles. The most visible ones are the swarms themselves, indexing sites and trackers.

Individual file or sets of files have their own unique swarm, an overlay network, in which peers trade pieces of file(s) referred to by a single .torrent. Dot-Torrent files are metadata files containing information that allows peers to check downloaded material for corruptions and to find other peers. Swarms are identified based on their infohash, the SHA1 hash of an immuteable part of the .torrent files. To join a swarm a user first has to find .torrent files he is interested in, download the .torrent and query a peer source for IP:Port lists to connect to.

Initially BitTorrent only supported trackers as peer sources. A tracker is a central servers managing peer lists for individual swarms. Such servers have to be included in the .torrent files at their creation or download time, and they are generally not updated by the users, even in the event of tracker failure.

Therefore trackers presented a single point of failure for swarms. Later, additional protocols to facilitate peer networking were introduced to mitigate this problem, the most important one is a Distributed Hash Table [2] based on the Kademlia [3] algorithm. The DHT is a decentralized overlay network in which most popular BitTorrent clients can participate in order store and retrieve peer lists for individual swarms.

Metadata files can be obtained in various ways, but most .torrents are downloaded from specialized indexing websites that list user-submitted .torrents, .torrents obtained from other websites or from a publisher of thematically related .torrents.

Two additional extensions [4][5] to the BitTorrent protocol also allow clients to obtain .torrent files via magnet-links from other peers which already have those .torrents. Magnet links are short URIs containing the infohash of a particular swarm and ,optionally, its tracker URLs. Combining these extensions it is possible to retrieve .torrent soley based on its infohash.

## 1.1 Motivation

At present there is no simple way for .torrent indexing websites to obtain complete lists of .torrents corresponding to currently active swarms. Torrent sites generally play a similar role to web search engines in the BitTorrent ecosystem. They usually rely on user-submitted content or scraping RSS feeds from other torrent sites, customized crawlers for original .torrent publishing sites and/or exchange of newly indexed .torrents between such indexing sites.

These approaches require manual work and may be complicated by linguistic differences between indexing sites and webmasters since there is no standard for .torrent exchange between websites. Therefore, even the biggest torrent sites only have an incomplete view of the available content.

The goal of this project is to aid the development of fully automated .torrent indexing services by obtaining .torrent files from the DHT in a sustainable manner, i.e. without placing an undue burden on the nodes in the DHT network.

## 1.2 Basic DHT operation

There actually are two Kademlia implementations for BitTorrent, one which is implemented by the Vuze client and one that became known as the Mainline DHT. Mainline DHT was originally created for the BitTorrent (Mainline) client and is now specified in BEP 05 [2]. Most methods presented in this paper are specific to the Mainline DHT as indexing the Mainline DHT was the aim of this work. The Vuze DHT implements some security measures to make indexing more difficult and generally requires a different approach as discussed by Wolchok et al. [6].

The Mainline DHT specification only describes the basic network protocol including the 4 RPC queries for basic kademlia operations. It is rather minimalistic since it allows *get*s and *put*s for IP:Port lists, i.e. it does not provide generic storage for binary data. It does not make use of enhanced Kademlia features such as replication and caching results along the path to the closest node sets.

Kademlia is based on an XOR distance metric that calculates node distances by XORing their IDs. With a bucketized routing table that is designed to have a near-exhaustive view of the keyspace close to the node's own ID while only having a random subset of far-away portions of the keyspace, the Kademlia-protocol guarantees both a lookup time complexity of $O(\log(n))$ and the routing table size scaling with $O(\log(n))$, where *n* is the number of nodes in the DHT network.

In the case of the BitTorrent DHT the keyspace is 160 bits wide which is the same length as the SHA1 hashes used for infohashes of the .torrent files.

For a node to announce itself it iteratively queries the closest nodes in its routing table for even closer nodes until no closer nodes are found and then asks those closest nodes to store its IP address and Port under the DHT key that is equal to the infohash.

A node that wishes to obtain peer lists for a specific torrent performs a similar lookup procedure but, instead of storing its own IP, it retrieves lists of IP-addresses and ports associated with that particular infohash.

BEncoding[1] is used in all queries used by the BitTorrent DHT as a network protocol. It is a simple encoding scheme for Maps, Lists, Integers and Strings that can be encoded and decoded in a canonical fashion. It was originally specified for the metadata file, but due to its flexibility and good forward-compatibility it has found use in many other aspects of BitTorrent clients. Since it is not a fixed binary header it cannot be be easily parsed unless the expected structure is known in advance. Intermediate representations, such as the primitive data structures corresponding to the bencoded data types provided by the used programming languages, are usually used before the data is actually processed. It is similar to JSON but does not require string escaping and is not meant to be human-friedly as it does not allow any ignoreable whitespaces, which could be used to prettify the raw data, although it still remains somewhat human-readable as long as the encoded values do not contain too much binary data.

---
1  BEncoding explained: http://wiki.theory.org/index.php?title=BitTorrentSpecification&oldid=3431#bencoding

More complex data types such as Objects are first decomposed into primitive types and then bencoded to conform with a specific structure of nested primitive data types as required by some protocol specification. I.e. BEncoding is usually used as an intermediate representation.

## 2 Design

The Kademlia protocol specified for BitTorrent use only supports two basic operations in relation to individual infohashes: `get_peers` and `announce_peer`. All other operations are solely for routing table maintenance. That means that neither enumerating available infohashes nor any other bulk lookups are supported.

Kademlia is also built on the assumption that all participating nodes are approximately equal, i.e. there are no supernodes, which generally are aware of more information in the network, as they can be found some other P2P overlays. Existing implementations may also perform filtering to prevent faulty or malicious nodes from congesting their links with requests or dominating the routing table of a node by spoofing their source addresses or frequently changing node IDs.

Therefore violating the protocol specification to increase the efficiency of the indexer would be of limited long-term use as client implementations might start defending against such violations. Placing excessive load on individual users would also discourage use of the DHT and therefore reduce the inclusiveness of the data collected.

To achieve efficient indexing despite these limitations the following synergetic optimizations were used:

### 2.1 Distributed Indexing

The most obvious approach to increasing throughput is using several DHT nodes instead of one. Using several ports on a single IP address was not considered a viable option due to IP-address based filtering against potential DoS attacks. Instead the indexer is designed to run on several hosts or on a multihomed host.

Individual instances synchronize their indexing activity through a shared relational database that stores discovered infohashes and the current processing stage for each .torrent file.

While IPv4 addresses are becoming more scarce every day it still should be possible to obtain a /24 subnet for indexing. Since an IPv6 extension for the BitTorrent DHT has already been specified[7] and implemented by some clients it is unlikely that address shortage will become a problem for indexing.

### 2.2 Passive Hash Collecting

All Kademlia lookups are iterative, meaning that they generally query distant nodes first and pick closer nodes from the resultant sets of previous steps in the lookup procedure. With a lookup concurrency of 10 – as used by this implementation – that means between 60 and 100 nodes are queried during an uncached lookup in the IPv4 DHT, which is estimated to have around 4M to 6M reachable nodes. Since each lookup request includes the target infohash any node visited during a lookup is able to learn about real infohashes.

The indexer makes use of this property by collecting all infohashes from incoming `get_peers` requests and incrementing a hit counter for already seen ones for garbage collection purposes. Other DHT data harvesting methods such as presented by Wolchok et al. [6][8] and Memon et al. [9] require nodes to be inserted in the closest-sets of the target keys or, if routing inaccuracies are taken into consideration, at least in their vincinity to intercept store or replication operations and collect their data. This means that only indexing nodes visited during the terminal phase of a storage-lookup produce useful data. With passive hash collecting useful data can be obtained at any point in the lookup, which greatly reduces the amount of nodes necessary to cover the entire keyspace. The downside of this method is that the indexer can only obtain the key-part of the stored key-value-pairs passively and has to perform active lookups to obtain the associated values. Essentially this trades peak throughput in key-value-pairs obtained per unit of time for a lowered minimal amount of nodes necessary for complete keyspace coverage.

This method is well-suited for continuous indexing since the structure of the routing table as defined by the Kademlia specification prefers long-lived nodes over short-lived ones to reduce the impact of churn on the network stability. Running the indexer on a server for extended periods of time increases the likelihood of being inserted into the root buckets of individual nodes and therefore improves the probability of being visited during lookups for infohashes that differ in the most significant bit from the querying node.

Running a long-lived node does not only increase the rate at which data is collected but also provides other nodes in the DHT with a stable reference point which presists across sessions for when reconnect to the DHT network at a later time.

### 2.2.1 Staggered Node IDs

When running a multihomed node each virtual node must assume a separate ID since several different IP:Port pairs using the same node ID is considered to be a violation of protocol constraints. Although using randomly selected IDs asymptotically yields good coverage of the DHT's keyspace better results can be achieved with a staggered distribution when only a few virtual nodes are running on one instance of the Indexer.

If for example the indexer would randomly choose the node IDs (in binary) 10111010... and 11101100... then it would only be inserted in the root bucket of nodes whose ID starts with a 0-bit while only ending up in buckets covering smaller ranges of the keyspace for all nodes starting with a 1-bit.

To achieve maximal coverage of the keyspace a random root node ID is chosen for an indexer instance and then a derived node ID is computed for each socket bound to each available IP on the host. The bits of the derived ID are calculated as $ID^{derived}_{160-i} = ID^{root}_{160-i} \oplus s_i$ based on the socket counter $s$. Using this algorithm each new socket is associated with a new node ID which has the maximal distance to all previously chosen IDs.

Better coverage of the ID space evens out biases in the distribution of infohashes seen in get_peers queries that are introduced by the layout of routing tables and also prevents the indexer from dominating any particular region of the key space. Therefore, it reduces the severity of any failure that can be caused by the indexer.

### 2.3 Shared routing table

The normal Kademlia routing table covers the 50% of the keyspace containing the keys differeing in the most significant bit with a fixed amount of nodes. The next-closer 25% of the keyspace are also covered by the same amount of nodes. This results in an individual node only having limited knowledge about the keyspace except for those parts which are close to the node's ID. While this is necessary to ensure a $O(\log(n))$ routing table size for a single node better results can be achieved if multiple nodes are running within the same instance.

Instead of using a separate routing table for each virtual node a shared routingtable for all nodes allows each of them to benefit of the keyspace information learned by the other ones. Combined with the node ID staggering this means that the routing table is more detailed over the entire keyspace. Depending on the number of virtual nodes used a shared routing table can reduce the remaining distance between the set of initially known IDs and the target ID by several bits.

Since the routing table is also used to respond to queries of external nodes the indexer also provides a good service to other nodes by providing better results than most normal nodes would.

## 2.4 Adaptive Timeouts

Since the DHT is intended to be run alongside with regular BitTorrent traffic, it can experience round trip times in the order of seconds or even completely dropped packets on heavily congested links when it has to compete with multiple TCP flows. As found by Crosby et al. [10] the baseline delay can vary from node to node, which requires conservative values to be chosen if fixed timeouts are to be used for queries. The default configuration was 10 seconds per call in the implementation on which the indexer was built. If all calls of a lookup are waiting for a timeout at the same time this can stall the entire lookup for several seconds.

In the absence of more complex solutions, such as the Vivaldi Coordinates [11] as they're used in the Bamboo [12] and Vuze DHTs to separately estimate timeouts for each query, self-tuning timeouts can yield significantly improved lookup performance and increase the overall throughput. This greatly reduces overall lookup time since the number of concurrently active lookups is limited to a fixed amount per virtual node in order to avoid excessive load on the DHT.

The timeout target is chosen as the round-trip time experienced by the 90$^{th}$ percentile of the last 256 successful queries. To avoid an asymptotic descent towards zero two separate timeouts are used, the adaptive timeout defined by the 90$^{th}$ percentile logic and a fixed timeout that is set to 10 seconds as before. The adaptive timeouts are based on calls considered successful under the fixed timeout while lookup procedures move on to the next nodes when the adaptive timeouts classify a query as stalled. Lookups still await the fixed timeout for the last pending queries to achieve maximum accuracy. This way stalled lookups are avoided while still being able to process the responses of all queries.

Each socket uses its own adaptive timeout estimation to account for potential differences in routing and between IPv4 and IPv6

## 2.5 Lookup Cache

To further reduce the amount of queries used per lookup a cache of nodes that have responded to recent requests is kept. This cache is shared among all lookups and updated in realtime. This allows concurrently running lookups to update their private candidate set of nodes-to-visit with node addresses from the shared cache since closer nodes may be discovered during the progress of other lookups.

Unlike the shared routing table which shortens the initial distance by a few bits for any target key the lookup cache shortens it by many bits but only for keyspace regions near target keys which have been recently visited. If a lookup is performed repeatedly on a single key or on a set of closely related keys this can reduce the usual 60-100 queries – for N = 10 – to 10 queries in the best case, i.e. achieve constant-time lookups instead of logarithmic-time ones.

## 2.6 In-order Traversal

Collected infohashes are initially stored in a database and not immediately queried. To achieve good performance the lookups should be performed in some optimized order instead. Most-frequently-seen and most-recently-added strategies were evaluated and initially the most-frequently-seen strategy proved to be the most useful. This latter strategy attempted to lookup infohashes that were likely to yield large peer lists and thus increased the chance of downloading a .torrent successfully. The effectiveness of this approach diminished as all low-hanging fruit were collected.

After simplifying the SQL-queries it turned out that fetching the infohashes in their natural order not only reduces the index size of the database table but also makes better use of the lookup cache.

Since there are approximately 6M reachable nodes in the BitTorrent DHT and the number of existing torrents is estimated to lie anywhere between 2M and 11M a significant overlap between the set of closest nodes responsible for individual infohashes can be expected. That is if they are adjacent to each other with regard to their natural ordering. Even in the case of non-overlap the nodes still should have more accurate information about nearby nodes, therefore drastically reducing lookup times. Fetching infohashes in their natural order ensures that most lookups are performed adjacent to recently-visited keys, which increases the likelihood of hits in the lookup cache.

It is important to note that the natural order only provides an approximation to closeness in the XOR metric since XOR does not account for carries. This can be easily seen in the following two examples:

A:  $011111010 + 000000100 = 011111110$
    $011111010 \oplus 011111110 = 000000100$

B:  $011111111 + 000000100 = 100000011$
    $011111111 \oplus 100000011 = 111111100$

While in *A* the natural and the XOR distance are equal they differ significantly in *B*. Such differences always occur on carries but the difference is small in most cases as the frequency of larger carries decreases exponentially with the number of bits affected. Therefore fetching .torrents in the natural order of their infohashes greatly increases the probability of a cache hit while keeping the complexity of database queries low.

To visualize the error introduced by this approximation Figure 1 below shows the distribution of natural distances of 5 million randomly generated keys sorted by their natural order when the distances are calculated between each key and its predecessor. Figure 2 uses the same naturally-ordered set of keys but shows the calculated xor-distance between each key and its predecessor in the natural order. At each integer power-of-two additional bits are carried and thus the distribution is spread to the higher distances, i.e. fewer shared prefix bits. It can be easily seen that the frequency of high-distance errors is low, e.g. the transition from the 0... to the 1... prefix only occurs once and thus only a single instance of neighbors having 0 bits in common can be found.

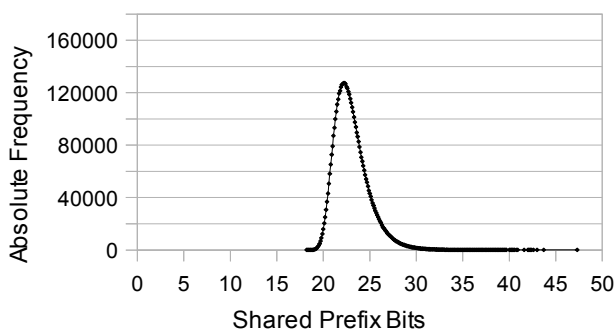 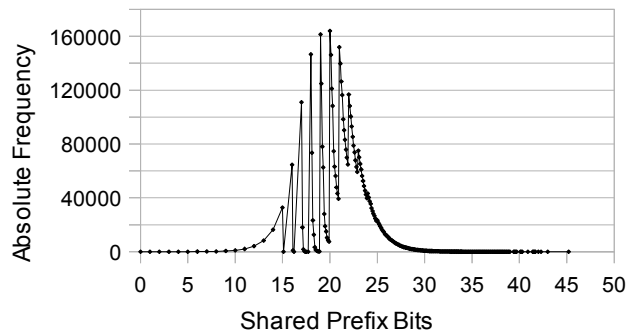

*Fig. 1: Distribution of natural distance between 5 million random keys*  *Fig. 2: Distribution of xor-based distance between the same keys*

### 2.6.1 Staggered Lookups

The simple in-order traversal strategy does not scale well when the number of virtual nodes increases since all nodes would try to query keys in the same keyspace range which would put a high load on a small, moving fraction of the DHT, i.e. it would only spread the load unevenly at any point in time. When that happens rate limiters of the affected nodes could lead to increased request drop rates and therefore decrease the efficiency of the indexer. To avoid this several in-order traversals are done in parallel with their starting points distributed using the same algorithm that is used to distribute the node IDs. This way the load on individual keyspace regions can be kept independent of the number of virtual nodes while the individual traversal tasks still benefit from the lookup cache.

## 2.7 Passive Retrieval

Inserting the indexer's IP address in all `get_peers` responses wouldn't be very scaleable since it would result in large amount of incoming connections and checking whether those hashes actually have to be retrieved in realtime would decrease the responsiveness of the DHT node as they would require a database query. But since many BitTorrent clients are behind NATs, which means the indexer cannot establish a connection to them, it is desireable to have the option to announce the indexer's presence and accept incoming connections as a fallback measure if active retrieval fails.

To achieve this a small table of recently-failed lookups is kept in memory and the indexer's address is only injected in `get_peers` responses that match the infohashes this table. This avoids altering responses unecessarily and significantly reduces the amount of incoming connections.

## 3 Implementation

The actual implementation of the DHT indexer[2] – written in Java – builds on the mlDHT[3] plugin for the BitTorrent client Vuze. The plugin's DHT implementation was designed for single-homed, low-throughput usage and had to be extended and rewritten significantly to fit the requirements of the indexer. The plugin also had no support for .torrent retrieval, it only acted as a peer source for the BitTorrent client.

### 3.1 Architecture

Due to its nature as a network application processing discrete UDP packets the DHT implementation itself is mostly event-driven and does all necessary processing in realtime on the IO-Threads, which means no long-running tasks may be started directly in response to incoming packets. Such long-running operations are separated into timed events that are handled by additional threads.

The indexer on the other hand can be conceptualized as a large state machine processing individual infohashes. Since the hashes have to be processed in bulk the actual implementation only uses the state concept in the database and otherwise acts as a pipeline of loosely coupled processing tasks. This is necessary since database queries may experience timeouts or at least take several seconds for larger queries and would block other aspects of the indexer if they weren't separated.

The individual steps of the infohash processing are performed by tasks executed periodically by a `ScheduledThreadPoolExecutor`. Handoff of Data Objects between those tasks is handled mostly through `ConcurrentLinkedQueues` and the database itself, although callbacks are used where the Indexer interacts with DHT and BitTorrent operations, i.e. lookups and metadata fetching.

---

2   sourcecode available under https://azsmrc.svn.sourceforge.net/svnroot/azsmrc/mldht/branches/indexer/
3   mlDHT Homepage: http://azureus.sourceforge.net/plugin_details.php?plugin=mlDHT

Infohashes generally are processed in the following order:

1. hashes from incoming `get_peers` requests are queued
2. duplicates in the queue are consolidated by a DB task and the hashes are inserted into the database or the hit counter for already existing hashes is incremented
3. infohashes without associated metadata are prefetched in their natural order from the database by a periodic task
4. DHT lookups are performed for the fetched infohashes until $3 \times number\ of\ sockets$ lookups are pending
5. finished lookups queue the discovered peer lists for further processing
6. peers are contacted through a minimalistic implementation of the metadata transfer protocol [5] using nonblocking IO
7. downloaded metadata is stored in the filesystem and database updates are again queued for the DB thread pool

Due to the high number of threads involved in all stages of processing the usage of `synchronized` primitives has been minimized in favor of the concurrency-friendly and often lock-free classes provided by the `java.util.concurrent` and `java.util.concurrent.atomic` packages. Most of these data structures rely on atomic compare-and-swap memory operations to achieve thread-safety without locking.

## 3.2 Database

Hibernate was chosen for the object-relational mapping library. It proved useful for the development process as it provides an abstraction over the underlying database and allowed the use of different database engines on the development and the target system. Namely HSQLDB was used for local development while MySQL's ACID-compliant InnoDB storage engine was used on the target server.

Initially simple row-locked read-modify-update database transactions were used. These seemed sufficient in the testing environment. However, on the target system with several dozen IP addresses it led to significant slowdowns after an hour of uptime. Profiling the application revealed that aborted transactions due to database-side deadlocks led to blocking waits and congestion of the database thread pool.

These deadlocks mostly occurred due to the random distribution of infohash updates falling into range-locks of other bulk queries and thus were inherent to the operation of the indexer.

After evaluating external applications that were using the same table it turned out that none of them used read-modify-write cycles and generally operated on rows with states that weren't used by the indexer. Under these conditions it was possible to lower the isolation level from *serializeable* to *repeatable read* combined with Hibernate's dynamic update query compilation that checks dirty fields of each row for modification in the `WHERE` clause of `UPDATE` statements, i.e. separate version or timestamp fields were not required to use multi-version concurrency control.

As mentioned earlier it turned out that fetching infohashes from the database in their natural order proved to be more efficient than other ordering strategies and reduced the amount of required indices significantly.

## 3.3 Routing table

To suit the needs of the indexer the original routing table implementation used by mlDHT had to be completely rewritten since it used a simplified approximation of the original Kademlia routing table design by keeping a list of buckets ordered by the number of leading bits shared with the local node ID.

When running on a multihomed host this approach is not viable since the node acts as several virtual nodes, each with its own node ID. Instead a modified version of the original Kademlia routing table was implemented. Instead of only splitting buckets when the bucket that is the prefix of the local node ID is full the modified version performs this split operation for any node ID used by the virtual nodes. This way all virtual nodes can share one routing table and due to the node ID derivation algorithm described above they have an even coverage of the keyspace, thus shortening any lookup by a few hops and providing better information to any external node that queries it.

Since the traffic of any of the virtual nodes is not localized to their respective local buckets any incoming query to any of the virtual nodes can affect any other portion of the routing table. Combined with the fact that buckets have to be informed of incoming traffic to update their replacement lists this results in heavily contended multi-threaded access to the routing table itself and the buckets within it.

Luckily Kademlia is designed to keep the content of the main buckets stable, i.e. read operations vastly outnumber the writes. Since using slightly outdated information during reads – e.g. while responding to `find_node` queries – is acceptable these characteristics lend themselves to a Copy-on-Write implementation for both the routing table and the individual buckets.

The routing table is implemented as a list of buckets associated with their prefixes, i.e. keys and the number of keyspace bits covered by a bucket. The list is sorted by said prefix to allow for fast lookups through binary search, i.e. expected $O(\log(b))$ performance where $b$ is the number of buckets which in turn scales with $O(k \cdot \log(n))$ with $k$ = number of virtual nodes and $n$ = number of nodes in the DHT.

Since bucket split and merge operations only occur rarely and modifying operations on the buckets themselves are checked against snapshots of the current state before a lock is acquired they are practically contention-free even with dozens of virtual nodes running on a single indexer instance.

Figure 3 is a simplified visualization of the shared IPv6 routing table for the 2 virtual nodes with the IDs F070E9... and 7070E9... The IPv4 table would have been significantly larger. The vertical position of each bucket represents the number of bits covered by its prefix, i.e. lower buckets cover a smaller fraction of the keyspace. The gray sub-buckets are the replacement lists. As it would also be expected for a single-nod routing table the buckets covering large segments of the keyspace are full while those close to the local IDs are not.

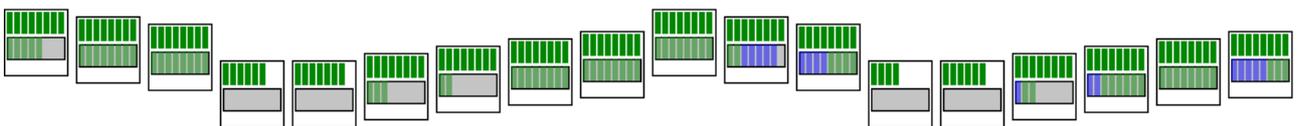

*Fig. 3: Visual representation of a shared routing table for 2 derived node IDs*

## 3.4 Shared Lookup Cache

The lookup cache is implemented as two sorted sets. The first set contains anchors: timestamped target keys of lookups for which the cache should keep addresses, and the second set contains the timestamped node addresses themselves, sorted by their node ID.

Similar to the infohash lookup strategy the cache uses the natural order of keys as an approximation to the XOR distance. Using this approximation it performs linear scans over a subset of the node-set to avoid inserting nodes into the cache if even-closer nodes are present, e.g. from lookups to nearby keys. Unlike the routing table which takes several failed requests to evict entries the cache immediately removes entries if the corresponding node fails to respond to a query. The failing node can be reinserted at a later point if it becomes responsive again.

Like the routing table, the cache is shared among all virtual nodes and all lookups, but unlike the routing table, it is under constant modification. Therefore using a Copy-on-Write strategy would be prohibitively expensive in terms of consumed CPU time. Instead Java's `ConcurrentSkipListMaps` are used which are lock-free implementations of a sorted set with similar performance characteristics to the commonly used `TreeMaps`. The major exception to this is that skip-lists only allow for linear-time in-order traversal while reverse order traversal has a $O(n \cdot \log(n))$ time complexity.

Additionally the anchor density can be greater than the node address density and vice versa. This complicates address removal routines as each node address might belong to several closest sets of different anchors. Therefore complete cleanups of the cache are only performed every 10 minutes while the checks performed during each insertion attempt may allow more nodes to be inserted into the cache than necessary. This is done to keep the complexity of insertions low at the expense of some additional memory use. In the worst case memory usage scales with $O(N \cdot a_T \cdot T)$ with $N$ = lookup concurrency, $a_T$ = number of anchors added during one cache timeout interval and $T$ = timeout interval for cache entries. In practice the overhead is lower since the anchor density is close to or even exceeds the node density if anchors are clustered due to the in-order traversal of the keyspace and therefore there is significant overlap between the closest-node-sets for each anchor, effectively reducing the number of anchors present.

## 3.5 Garbage Collector optimizations

After initial optimizations yielded acceptable throughput, profiling the indexer revealed that not only the code execution itself incurred significant costs, but that the creation of many short-lived objects caused excessive GC churn. This sometimes caused several minor garbage collections per second and on a few occasions even caused `OutOfMemory` exceptions due to the Garbage Collector not meeting its throughput goal[4].

The most significant cause of GC churn was UDP packet decoding and encoding. The specification for the BitTorrent DHT uses bencoding to serialize the parameters and values of each request and response. As mentioned earlier Bencoding is a binary format to represent nested primitive data types such as strings, integers, lists and maps. mlDHT and therefore the indexer make use of the Vuze `BEncoder/BDecoder` classes to transform UDP packets into primitive java types, which are then again converted into DHT-specific types by a separate decoder. While most decoding and encoding steps were already optimized for either low CPU usage or low long-term memory footprint, several steps involved in the process unnecessarily create temporary objects, i.e. short-term memory usage, which can lead to excessive GC churn when the encoders/decoders are used in a high throughput scenario.

By improving these encoders, and submitting patches to the Vuze source repository, and several smaller hotspots the object churn was significantly reduced. Tweaking the garbage collector parameters of the Sun JVM to increase the size for the young generation – the memory space used for new/short-lived objects – increases the amount and fraction of objects that can be discarded with a single minor collection and thus reduces the number of minor collections needed.

Overall these changes lead to less time spent in garbage collection which also increases the throughput as the worker threads have to be suspended less often.

---

4  http://java.sun.com/javase/technologies/hotspot/gc/gc_tuning_6.html#par_gc.oom

## 3.6 Selective Queue Management

Since individual tasks of the indexer are only coupled through queues some network-facing components may produce more events than can be processed, which means some of them must be evicted from the queue without processing. For example the rate of incoming `get_peers` queries may exceed the throughput that the database backend can achieve. Instead of just randomly dropping queue entries, as some active queue management (AQM) algorithms do, an adaptive replacement cache (ARC) [13] was used to build blacklists of infohashes that should not be processed again soon. This allows the indexer to mostly evict recently-seen or frequently-seen infohashes from its processing queue to process the rarer infohashes instead, which prevents popular torrents from drowning out the signal of less popular ones.

Since the hit rate of an ARC-based filter depends on the popularity of individual infohashes is distributed and the size of the filter itself no static filter would manage to limit the throughput to a specific goal. A modified version was used instead that could be resized based on queue over- and underflows. Essentially the ARC-filter was combined with the Blue [14] algorithm to scale it to different throughput scenarios without any manual parameter tuning.

The ARC+Blue filter was also used as whitelist for infohashes in the passive retrieval stage to dynamically limit the number of modified responses and thus the amount of incoming connections.

## 4 Performance evaluation

To better emulate steady-state operation of the indexer, the database on the target system was pre-filled with 5 million hashes of already-known .torrents and currently-unknown infohashes provided by administrators of two large torrent indexing sites. This reduced the number of low-hanging fruit, i.e. large pre-existing swarms to be indexed.

The latest version was tested on a 2 GHz 4-core AMD Opteron Server with 4GB RAM running a 64bit Linux system including a 64bit Sun Java 1.6 Server VM.

Internally the number of concurrent lookups was restricted to $3 \cdot min(IPv4\ addresses, IPv6\ addresses)$, with 65 IPv6 and 64 IPv4 addresses routed to the server this effectively allowed 192 concurrent concurrent lookups to be performed at any time.

After one week of uptime the indexer had settled into a combined Rx/Tx rate of 2,500 to 3,000 packets per second and 4,800 to 5,000 kbit/s. Split across the virtual nodes that is approximately 10 to 20 times the load that the same amount of normal DHT nodes would cause. During that week the indexer obtained 587,000 new .torrents and is still collecting more than 330 new .torrents every 10 minutes. Still, this is just a lower bound since infohashes which are deleted from the database due to repeatedly failed lookups not being included in this measurement.

According to profiling information the indexer's memory usage ranges between 150 and 230MB worth of objects on the java heap, depending on various internal cleanup tasks. CPU usage is constant most of the time and hovers around 60% (out of a maximum of 400% on the quad-core system) and occasionally increases the load to 100% on one core when long-running tasks are executed.

When fed with a dataset containing many adjacent infohashes obtained from several tracker servers the indexer performed up to 4600 lookups in a 10 minute interval, which yielded up to 870 .torrent files in the same interval.

## 5 Conclusion

Overall the DHT indexer can obtain new torrents faster than most existing indexing services, is fully automated and can run with moderate hardware requirements. Running it in a distributed manner spread over several hosts could further decrease the load on individual hosts. Thus the indexer could perform the same task using the spare processing power and memory of multiple hosts.

Probably the most significant hurdle to operating an indexer right now is obtaining a sufficient amount of IP addresses. This problem should become less significant in the future as IPv6 adoption increases.

This demonstrates the feasibility of efficient, decentralized indexing without violating protocol constraints and can be used to increase the resilience of the BitTorrent ecosystem by allowing new indexing sites to be built without significant manual work.

Fully automated indexing may inadvertently violate the privacy of some BitTorrent users since it obtains all torrents that are tracked through the DHT, even if the torrent creator did not intend to publish them to any website. Although the private flag [15] is supposed to prevent such leakage, it relies of conformance of all BitTorrent clients and on the user's knowledge to mark torrents as private. Since BitTorrent is designed as an open system such problems would have to be resolved at a different level, for example via transparent encryption of transferred content or simply by educating users.

## 6  Future Work

While obtaining new torrents is one important aspect for indexing websites they also fulfill other roles such as verifying a torrents authenticity and providing statistics indicating the torrent's availability. The latter is usually based on client counts obtained from tracker servers. Although a DHT extension to obtain such statistics has already been specified [16] it is not widespread yet. Once it is adopted by the majority of bittorent clients further improvements in instrumenting the DHT may be required to gather such statistics in bulk without disrupting the DHT and without investing a large amount of resources.

Algorithm" , *IEEE Computer*, April 2004

14: W. Feng, D. Kandlur, D. Saha, K. Shin, Blue: A New Class of Active Queue Management Algorithms, U. Michigan, April 1999

15: David Harrison, "Private Torrents", 2008, http://bittorrent.org/beps/bep_0027.html

16: Aaron Grunthal, "DHT Scrapes", 2010, http://bittorrent.org/beps/bep_0033.html